\documentclass[prd,twocolumn,amsmath,amssymb]{revtex4}
\usepackage[colorlinks,linkcolor=blue,anchorcolor=blue,citecolor=blue,urlcolor=blue]{hyperref}
\usepackage{graphicx}
\usepackage{dcolumn}
\usepackage{bm}
\usepackage{overpic}
\usepackage{rotating}
\usepackage{epstopdf}
\usepackage{color}
\usepackage{verbatim}

\newcommand{\jpsi}{J/\psi}
\newcommand{\EE}{e^+e^-}
\newcommand{\MM}{\mu^+\mu^-}
\newcommand{\TT}{\tau^+\tau^-}
\newcommand{\Bee}{\mathcal{B}_{e^+e^-}}
\newcommand{\Bmm}{\mathcal{B}_{\mu^+\mu^-}}
\newcommand{\Btt}{\mathcal{B}_{\tau^+\tau^-}}
\newcommand{\Bll}{\mathcal{B}_{l^+l^-}}
\newcommand{\GG}{\gamma\gamma}

\def\babar{\mbox{\slshape B\kern-0.1em{\small A}\kern-0.1em B\kern-0.1em{\small A\kern-0.2em R}}}
\begin{document}
\normalsize
\parskip=5pt plus 1pt minus 1pt
\title{Precision measurement of the branching fraction for the decay $\psi(2S)\rightarrow\tau^{+}\tau^{-}$}
\author{
 M.~Ablikim$^{1}$, M.~N.~Achasov$^{4,c}$, P.~Adlarson$^{76}$, O.~Afedulidis$^{3}$, X.~C.~Ai$^{81}$, R.~Aliberti$^{35}$, A.~Amoroso$^{75A,75C}$, Y.~Bai$^{57}$, O.~Bakina$^{36}$, I.~Balossino$^{29A}$
% ...
\\
\vspace{0.2cm}
(BESIII Collaboration)\\
\vspace{0.2cm} {\it
$^{1}$ Institute of High Energy Physics, Beijing 100049, People's Republic of China\\
% ...
$^{3}$ Bochum  Ruhr-University, D-44780 Bochum, Germany\\
$^{4}$ Budker Institute of Nuclear Physics SB RAS (BINP), Novosibirsk 630090, Russia\\
% ...
$^{29}$ INFN Sezione di Ferrara, (A)INFN Sezione di Ferrara, I-44122, Ferrara, Italy; (B)University of Ferrara,  I-44122, Ferrara, Italy\\
% ...
$^{35}$ Johannes Gutenberg University of Mainz, Johann-Joachim-Becher-Weg 45, D-55099 Mainz, Germany\\
$^{36}$ Joint Institute for Nuclear Research, 141980 Dubna, Moscow region, Russia\\
% ...
$^{57}$ Southeast University, Nanjing 211100, People's Republic of China\\
% ...
$^{75}$ University of Turin and INFN, (A)University of Turin, I-10125, Turin, Italy; (B)University of Eastern Piedmont, I-15121, Alessandria, Italy; (C)INFN, I-10125, Turin, Italy\\
$^{76}$ Uppsala University, Box 516, SE-75120 Uppsala, Sweden\\
% ...
$^{81}$ Zhengzhou University, Zhengzhou 450001, People's Republic of China\\
\vspace{0.2cm}
% ...
$^{c}$ Also at the Novosibirsk State University, Novosibirsk, 630090, Russia\\
% ...
}
}
\date{\today}
\begin{abstract}
	{Using $(2259.3 \pm 11.1)\times10^{6}$ $\psi(2S)$ events acquired with the BESIII detector, the branching fraction of $\psi(2S)\rightarrow\tau^{+}\tau^{-}$ is measured with improved precision to be $\mathcal{B}_{\psi(2S)\rightarrow\tau^{+}\tau^{-}}=(3.240~\pm~0.023~\pm~0.081)\times 10^{-3}$, where the first and second uncertainties are statistical and systematic, respectively, which is consistent with the world average value within one standard deviation. This value, along with those for the branching fractions of the $\psi(2S)$ decaying into $e^{+}e^{-}$ and $\mu^{+}\mu^{-}$, is in good agreement with the relation predicted by the sequential lepton hypothesis. Combining the branching fraction values with the leptonic width of the $\psi(2S)$, the total width of the $\psi(2S)$ is determined to be (287 $\pm$ 9) keV. }
\end{abstract}
\maketitle
\section{\boldmath Introduction}
{ In recent years, experimental studies of $B$ meson decays have hinted toward deviations from  the Standard Model (SM) expectations of Lepton Flavor Universality (LFU) in semi-leptonic decays~\cite{BaBar:2012obs,BaBar:2013mob,Belle:2015qfa,Belle:2016ure,LHCb:2015gmp}. Many experiments have provided evidence suggesting that it may be better to probe LFU by searching for deviations of the ratios
\begin{eqnarray}
R(D^{(*)})=\frac{\Gamma(B\rightarrow D^{(*)}\tau^+\nu_\tau)}{\Gamma(B\rightarrow D^{(*)} l^+\nu_l)} , (l= e, \mu).
\end{eqnarray}
from their expected values. The combined results of BaBar~\cite{BaBar:2012obs,BaBar:2013mob}, Belle~\cite{Belle:2015qfa,Belle:2016ure} and LHCb~\cite{LHCb:2015gmp} show a deviation from the SM prediction by 3.1$\sigma$~\cite{Martinelli:2021ccm}.
The $R(D^{(*)})$ puzzle has garnered significant attention in recent literature. Various studies have interpreted this deviation using effective field theory, and developed explicit new physics (NP) models.
~A precise measurement of the process $c\bar{c}\rightarrow \TT$ can therefore serve as a benchmark test for new theoretical models~\cite{Aloni:2017eny}.
Meanwhile, this process also affords a unique platform to compare the three generations of the leptonic weak interaction by studying the leptonic decays of
$\psi(2S) \to \EE~,~\MM$,~and $\TT$. The sequential lepton hypothesis leads to a relationship between the branching fractions of these decays, $\Bee$, $\Bmm$, and $\Btt$, given by
\begin{eqnarray}
\frac{\Bee}{v_e (\frac{3}{2}-\frac{1}{2} v_e^2)}=
\frac{\Bmm}{v_\mu (\frac{3}{2}-\frac{1}{2} v_\mu^2)}=
\frac{\Btt}{v_\tau (\frac{3}{2}-\frac{1}{2} v_\tau^2)},
\label{Eq.fracrelation}
\end{eqnarray}
with $v_l=[1-(4m_l^2/M^2_{\psi(2S)})]^{1/2}$ and $l=e,\mu,\tau$. Substituting the nominal masses of the leptons and $\psi(2S)$, we obtain
\begin{eqnarray}
\Bee\simeq\Bmm\simeq\frac{\Btt}{0.3890}\equiv \Bll~~.
\label{Eq.fracthree}
\end{eqnarray}
The BES Collaboration preformed such a comparison using 3.96 million $\psi(2S)$ events in 2002~\cite{BES:2000wyf}. A subsequent measurement using a similar method improved upon this initial study in 2006~\cite{BES:2006jil}. Lately, the BESIII Collaboration collected about 2259 million $\psi(2S)$ events in 2021~\cite{BESIII:2023psipError}. Such a large data sample allows for a more precise study. In addition, a precise measurement of $\Btt$
can help us better determine the total number of $\tau$ pairs produced, which can then be used to search for and study $\tau$-related decays.

This paper reports a precise measurement of $\Btt$ using BESIII data mentioned above, and it is organized
as follows. Sec.~\ref{DTmcSample} contains a brief description of the BESIII detector, datasets, and Monte Carlo (MC) samples.
In Sec.~\ref{EvtBkg}, the event selection and the background estimation are described. In Sec.~\ref{CrsCal}, we describe the cross section calculation of the continuum process $\EE\rightarrow\TT$, and in Sec.~\ref{BFrCal} the measurement of $\mathcal{B}_{\tau^+\tau^-}$ is presented. In Sec.~\ref{Uncertainty}, the systematic uncertainties are discussed. The conclusions are summarized in Sec.~\ref{Summary}.
}
\section{\boldmath Detector and data samples}\label{DTmcSample}
{
The BESIII detector~\cite{Ablikim:2009aa} records symmetric $e^+e^-$ collisions provided by the BEPCII storage ring~\cite{Yu:IPAC2016-TUYA01} in the center of mass energy range from 1.84 to 4.95 GeV, with a peak luminosity of
1.1$\times$$10^{33}$ cm$^{-2}$s$^{-1}$ achived at $\sqrt{s}$ = 3.773 GeV. BESIII has collected large data samples in this energy region~\cite{Ablikim:2019hff}.
The cylindrical core of the BESIII detector covers $93\%$ of the full solid angle and consists of a helium-based multilayer drift chamber~(MDC), a plastic scintillator time-of-flight system~(TOF), and a CsI(Tl) electromagnetic calorimeter~(EMC), which are all enclosed in a superconducting solenoidal magnet providing a magnetic field of about 1.0~T. The solenoid is supported by an octagonal flux-return yoke with resistive plate counter muon identification modules interleaved with steel. The charged-particle momentum resolution at $1~{\rm GeV}/c$ is $0.5\%$, and the d$E$/d$x$ resolution is $6\%$ for electrons from Bhabha scattering. The EMC measures photon energies with a resolution of $2.5\%$ ($5\%$) at $1$~GeV in the barrel (end-cap) region. The time resolution in the TOF barrel region is 68~ps, while that in the end-cap region is 110~ps. The end-cap TOF system was upgraded in 2015 using multi-gap resistive plate chamber technology, providing a time resolution of 60~ps~\cite{etof1,etof2,etof3}.

Simulated MC samples of signal and background processes are produced to optimize the event selection criteria, determine the detection efficiency and estimate the background contamination. The response of the detector is reproduced using a {\sc geant4}-based~\cite{GEANT4} {MC} simulation software package, which includes the geometric and material description of the BESIII detector, the detector response and digitization models.

In the inclusive MC sample, the production of the $\psi(2S)$ resonance is simulated by the \mbox{KKMC} generator~\cite{Jadach:2000ir}. Known decay modes are generated using EVTGEN~\cite{Ping:2008zz} with the branching fractions set at world-averaged values~\cite{PDG2022}, while the remaining unknown decay modes are modeled by \mbox{LUNDCHARM}~\cite{Chen:2000tv}. Final state radiation (FSR) from charged final state particles is incorporated using the PHOTOS package~\cite{Richter-Was:1992hxq}. As for the quantum electrodynamics (QED) background, the $e^{+}e^{-}\rightarrow e^{+}e^{-}$ (Bhabha),  $e^{+}e^{-}\rightarrow \mu^{+}\mu^{-}$ and $e^{+}e^{-}\rightarrow \GG$ processes are generated by BABAYAGA~\cite{BABAYAGA}. The two-photon processes $e^{+}e^{-}\rightarrow e^{+}e^{-} X$ with $X=e^{+}e^{-}, \mu^{+}\mu^{-}, \eta, \eta^{\prime}, \pi^{+}\pi^{-}$, and $K^{+}K^{-}$~\cite{TwoPhoton} are simulated using the generators DIAG36~\cite{DIAG36}, EKHARA~\cite{EKHARA}, and GALUGA2.0~\cite{GALUGA2.0}, respectively.

The signal MC samples for the process \mbox{$\psi(2S) \rightarrow \tau^+ \tau^-$} ($\tau^{+}\rightarrow \mu^{+}\nu_{\mu}\bar{\nu_{\tau}}$, \mbox{$\tau^{-}\rightarrow e^{-}\bar{\nu_{e}}\nu_{\tau}$} or $\tau^{+}\rightarrow e^{+}\nu_{e}\bar{\nu_{\tau}}$, \mbox{$\tau^{-}\rightarrow \mu^{-}\bar{\nu_{\mu}}\nu_{\tau}$}) are generated using the ConExc~\cite{ConExc2014} generator with the model PHOTOS, and the subsequent decays of the $\tau$ are modeled by EVTGEN~\cite{tauleptonnunu} with the model TAULNUNU. In order to account for the effect of initial state radiation (ISR) on the detection efficiencies, the lineshape of the Born cross section for $e^+e^-$ continuum processes, obtained from calculations at energy points
from 3.670 GeV to 3.700 GeV in the vicinity of the $\psi(2S)$, and considering the effect of the beam energy spread (Section~\ref{CrsCal}), is used as input into the ConExc generator to simulate the signal processes.

\section{\boldmath Event Selection and Background Analysis}\label{EvtBkg}
{
 Candidate events for $\psi(2S)\rightarrow \tau^+\tau^-$ are identified using pure leptonic decay modes of the $\tau$, with one $\tau$ decaying into an electron and the other into a muon final state. The accompanying neutrinos cannot be detected by the BESIII detector. Therefore, the final state of interest has two charged tracks. We select candidate events with a total number of charged tracks equal to two and net charge equal to zero. Charged tracks
are reconstructed using the MDC information. For each track, the point of closest approach to the interaction point (IP) must be within 1 cm in the plane perpendicular to the beam direction and within 10 cm along the beam direction. Moreover they should be in a polar angle range $|\cos\theta|<$0.93, where $\theta$ is the angle between the direction of the track and the beam direction ($z$ axis) to guarantee the best agreement between data and MC simulation. Furthermore, a vertex fit, in which the tracks are forced to pass through a common vertex, is performed and required to be successful. The momentum of each track is required to be less than 1.2 GeV$/c$ based on signal MC studies.

Photons are identified as showers within the EMC. The deposited energy of each shower must be greater than
25 MeV in the barrel region ($|\cos\theta|<0.80$) and greater than 50 MeV in the end cap region ($0.86< |\cos\theta|<0.92$). To exclude showers originating from charged tracks, the angle between the position of each shower in the EMC and the closest extrapolated charged track must be greater than 10 degrees.
To suppress electronic noise and showers unrelated to the event, the difference between the EMC time and the event start time is required to be within [0, 700] ns. The number of photons in the EMC is required to be zero.

\begin{figure}[htbp]
		\begin{overpic}[width=0.47\textwidth]{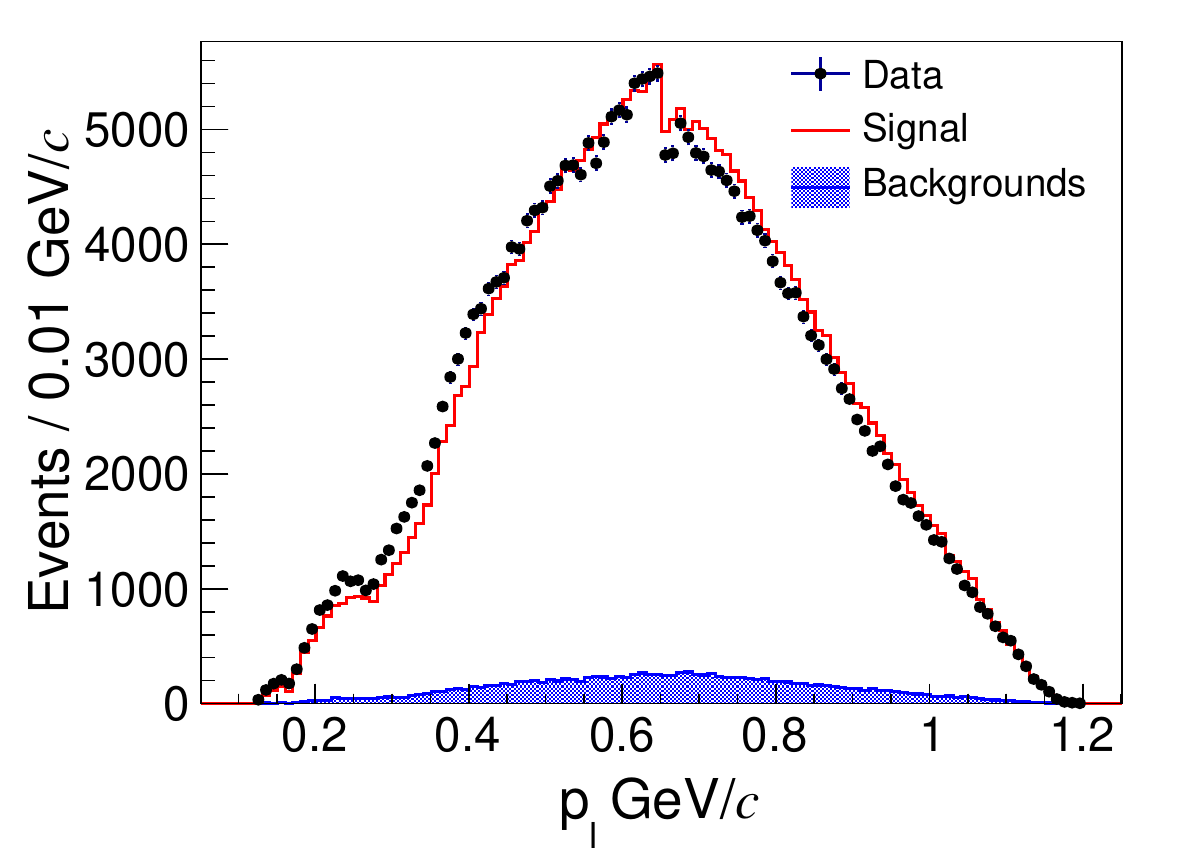}
        \end{overpic}
		\begin{overpic}[width=0.47\textwidth]{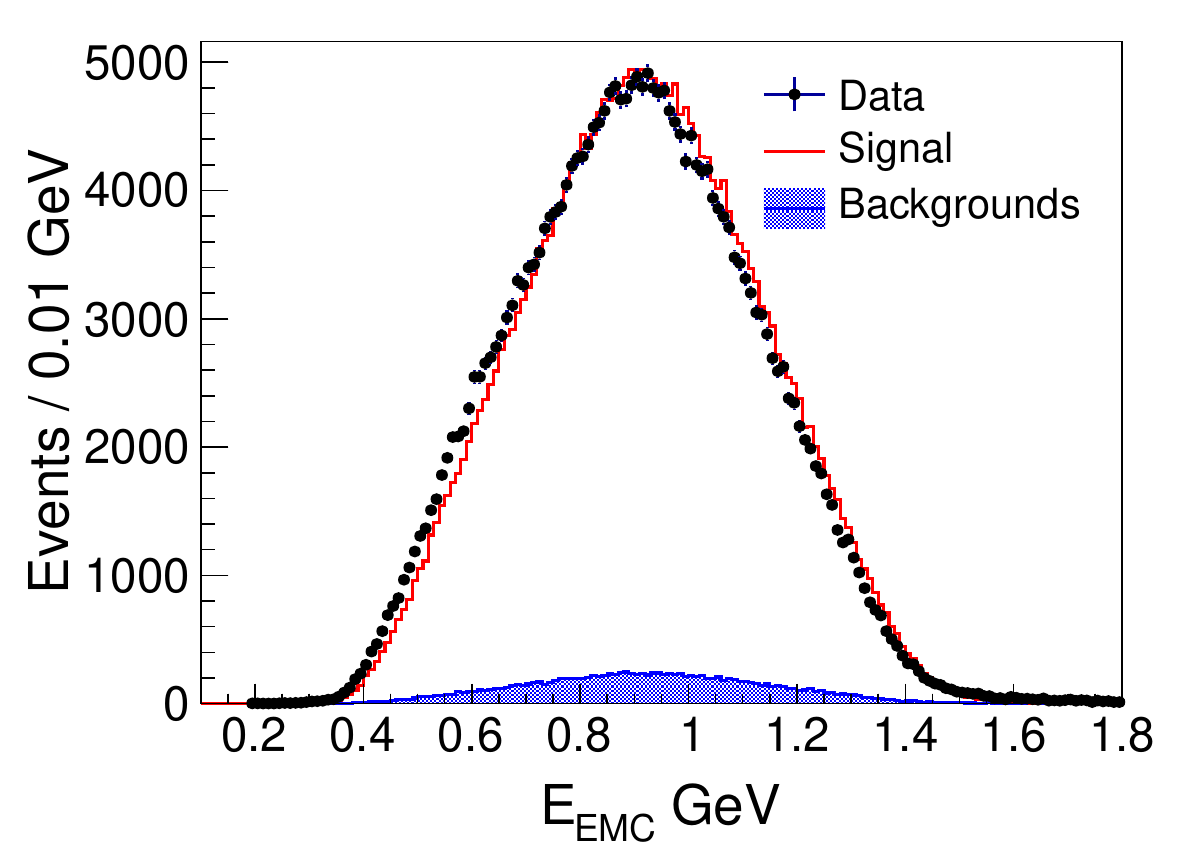}
        \end{overpic}
\caption{Distribution of the lepton momentum~(top) and the total energy deposited in the EMC by the leptons~(bottom). The dots with error bars are data; the red histograms are the signal from MC simulation; and the blue shaded histograms are background from the inclusive MC sample.}
\label{figure1:momPanddepE}
\end{figure}

Furthermore, the two charged tracks are required to be an electron and a muon, respectively.
A charged track is identified as an electron if it satisfies the following requirements: the ratio of deposited energy in the EMC to its momentum ($E/p(e)$) falls between 0.8 and 1.2, the $\chi^{2}_{dE/dx}(e)$ of the fitted track in the MDC is less than 4, and the difference between the expected flight time and the measured time in the TOF $|\Delta tof|(e)$ is less than 0.3 ns. The other charged track is identified as a muon if $E/p(\mu)<0.7$,
$\chi^{2}_{dE/dx}(\mu)<4$ and  $|\Delta tof|(\mu)<0.3$~ns.
To distinguish between muons and pions, we impose an additional requirement on the depth of the track in the MUC. This depth must be greater than $a\times(p-b)$~cm, where the parameters $a=$ 81 cm and $b$ = 0.65 GeV/$c$ are derived by optimizing a figure of merit~(FOM) related to a two-dimensional plot of the depth length
in the MUC versus the track momentum $(p)$. The FOM is defined as ${N_{\rm sig}}/\sqrt{N_{\rm sig}+N_{\rm bkg}}$, where $N_{\rm sig}$ and $N_{\rm bkg}$ are the numbers of signal and background events estimated by the inclusive MC samples, respectively.
In order to distinguish the signal and background events, the following variables are used:
\begin{eqnarray}
P4_{\rm mis}=P4_{\psi(2S)}-P4_{e\mu},
\end{eqnarray}
\begin{eqnarray}
\cos\theta_{\rm mis}=\frac{-(p_{e}+p_{\mu})_{z}}{|p_{e}+p_{\mu}|},
\label{equation:polarangle}
\end{eqnarray}
\begin{eqnarray}
U_{\rm mis}=E_{\rm mis}-P_{\rm mis},
\label{equation:umis}
\end{eqnarray}
where $P4_{\rm mis}$, $P4_{\psi(2S)}$, and $P4_{e\mu}$ are the four momenta of the missing particles, the $\psi(2S)$ resonance and the leptonic pairs (electron-muon pairs), respectively.
The $M_{\rm mis}$ and $\theta_{\rm mis}$ variables denote the invariant mass, calculated using the missing four momentum ($P4_{\rm mis}$), and the polar angle of the missing particles, calculated according to Eq.~\ref{equation:polarangle}.
Finally, we require that the missing mass within each event,~$M_{\rm mis}$, must be less than 3.05 GeV$/c^{2}$, which is obtained by optimization and removes $\pi^{+}\pi^{-}J/\psi$ backgrounds. To reject two-photon backgrounds, the requirement $|\cos\theta_{\rm mis}|<0.8$ is used.
After all selection criteria are applied, 280,412 candidate events remain in data.
The distributions of lepton momentum (p$_{l}$) and the total energy deposited by the two leptons in the EMC ($\rm {E_{EMC}}$) for these candidate events are shown in Fig.~\ref{figure1:momPanddepE}.

Potential backgrounds are estimated using an inclusive MC sample of $\psi(2S)$ events, as detailed in Sec.~\mbox{\ref{DTmcSample}}. The background study is performed relying on MC-Truth matching and utilizing the TopoAna package~\cite{Zhou:2020ksj}. After all the selection criteria are applied, the missing mass and $U_{\rm mis}$ distributions are shown in Fig.~\ref{figure2:misMandmisU} for data, the signal MC sample, and the residual background components.
The dominant backgrounds are from the decays \mbox{$\psi(2S)\rightarrow \TT\rightarrow e^{+}\nu_{e}\bar{\nu_{\tau}}\pi^{-}\bar{\nu_{\tau}}$} and $\psi(2S)\rightarrow\TT\rightarrow \pi^{+}\bar{\nu_{\tau}}e^{-}\bar{\nu_{e}}\nu_{\tau}$. The number of background events is estimated to be 11,531 based on the inclusive MC sample.
\begin{figure}[htbp]
	\begin{center}
		\begin{overpic}[width=0.47\textwidth]{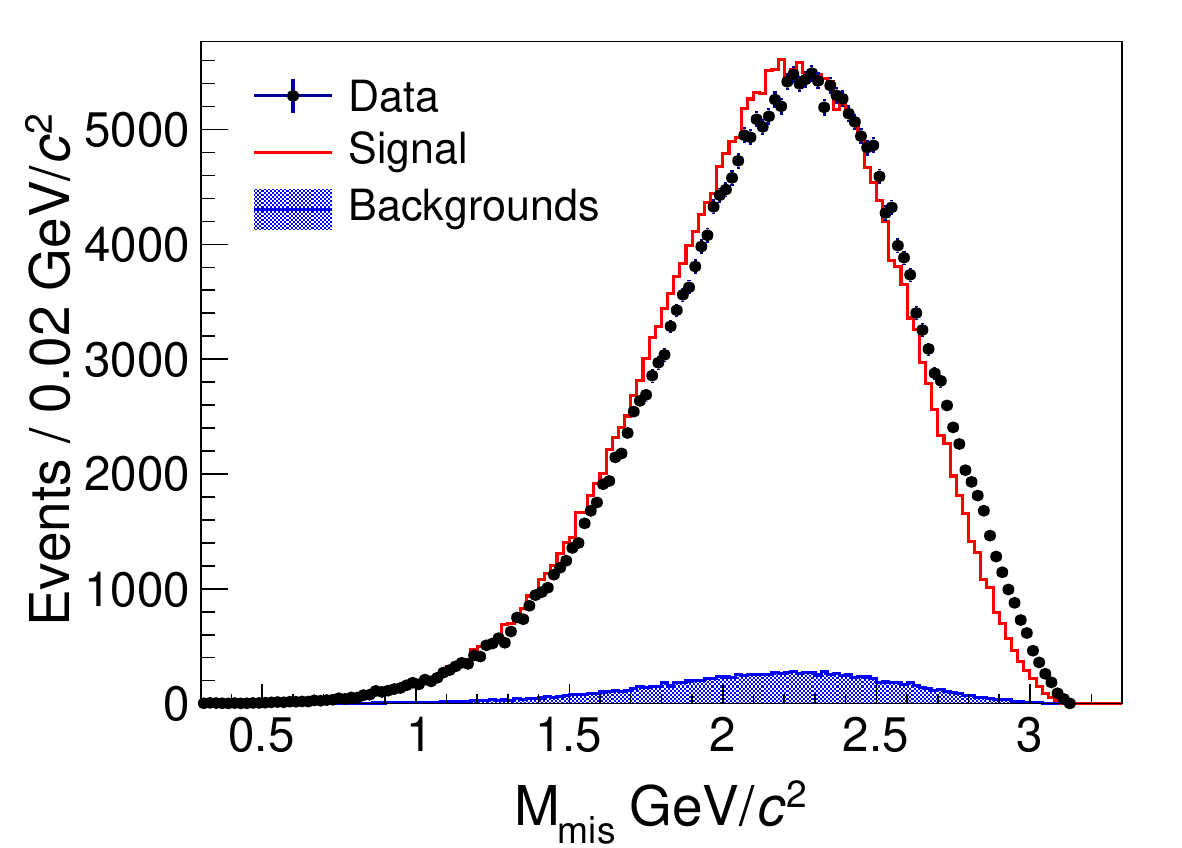}
        \put(0.2,5000){(a)}
		\end{overpic}
		\begin{overpic}[width=0.47\textwidth]{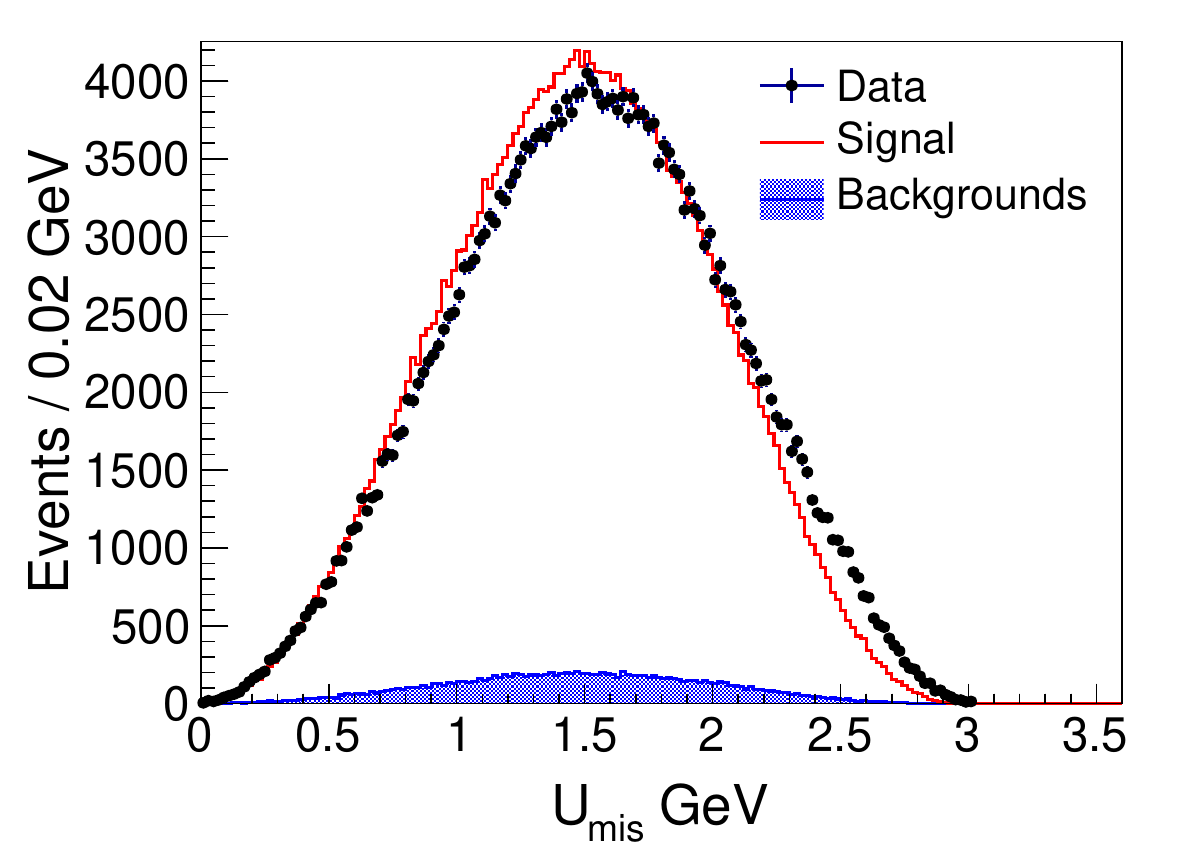}
        \put(-450,250){(b)}
		\end{overpic}
	\end{center}
	\caption{Distribution of the variable M$_{\rm mis}$ (top) and U$_{\rm mis}$ (bottom). The dots with error bars are data; the red histograms are the signal from MC simulation; and the blue shaded histograms are background from the inclusive MC sample.}
	\label{figure2:misMandmisU}
\end{figure}
  The backgrounds from two-photon and QED processes are estimated using exclusive MC samples. The number of background events from two-photon processes $e^{+}e^{-}\rightarrow e^{+}e^{-} X$ ($X=e^{+}e^{-}, \mu^{+}\mu^{-}, \eta, \eta^{\prime}, \pi^{+}\pi^{-}$, and $K^{+}K^{-}$) is about 308, while the effect of QED backgrounds (Bhabha, di-$\mu$ and di-$\gamma$) is negligible. The total number of background events is estimated to be 11,839, with a fraction of approximately 4.3$\%$.
}
\section{\boldmath Cross section calculation on $\EE\rightarrow\TT$}\label{CrsCal}
{
Experimentally, the measured total $\tau$-pair production cross section has the form~\cite{Kuraev:1985hb}:
\begin{eqnarray}
\begin{split}
\sigma_{\rm exp}(s,m_{\tau},\Delta)=\int_{0}^{\infty}
d\sqrt{s'} G(\sqrt{s'},\sqrt{s})\\
\times \int_{0}^{ 1-\frac{ 4m_{\tau}^2 }{ s' } } dx F(x,s')
\frac{ \bar{\sigma}_{\rm tot} (s'(1-x),m_{\tau}) }{|1-\Pi (s'(1-x))|^2}~.
\end{split}
\label{expsec}
\end{eqnarray}
Here, $F(x,s)$ is the ISR factor~\cite{Kuraev:1985hb}, parameterized as
\begin{eqnarray}
\begin{split}
F(x,s)&=\beta x^{\beta-1} [1+\frac{3}{4}\beta+
   \frac{\alpha}{\pi} \frac{\pi^2}{2}-\frac{1}{2}
   +{\beta^2}\frac{9}{32}\\
   & -\frac{1}{12}\pi^2 ]
     -\beta (1-\frac{1}{2} x )
       + \frac{1}{8}\beta^2
       [ 4(2-x)\\
   & \ln\frac{1}{x}
     -\frac{1+3(1-x^2)}{x}
     \ln (1-x)-6+x ] ,
\end{split}
\label{rdnftr}
\end{eqnarray}
with
$${\displaystyle \beta=\frac{2\alpha}{\pi} \left(\ln \frac{s}{m^2_e}-1 \right).~~}$$
$\Pi$ is the vacuum polarization factor~\cite{Ruiz-Femenia:2001qlg, Berends:1973tz, Berends:1976zn}, and
$G(\sqrt{s'},\sqrt{s})$, is usually treated as a Gaussian distribution~\cite{Wang:2002np}
$$G(\sqrt{s'},\sqrt{s})=\frac{1}{2\pi\Delta}
e^{\frac{-(\sqrt{s}-\sqrt{s'})^2}{2\Delta}}~~,$$
where $\Delta$ depicts the energy spread of the $e^+e^-$ collider. We explicitly indicate the dependence of the
cross section on $s$ and $m_\tau$ in Eq.~(\ref{expsec}).

The total cross section $\bar{\sigma}_{\rm tot}$ consists of three parts,
\begin{eqnarray}
\bar{\sigma}_{\rm tot} (s) =F_s(s)\cdot (\sigma_{\rm con}(s) + \sigma_{\rm int}(s) +\sigma_{\rm res}(s))/\sigma_{\rm con}(s)~~,
\label{totsxn}
\end{eqnarray}
where $\sigma_{\rm con}(s)$, $\sigma_{\rm res}(s)$, and $\sigma_{\rm int}(s)$ stand for the continuum, resonance, and interference parts, respectively, defined as
\begin{eqnarray}
\begin{array}{rcl}
\sigma_{\rm con}(s)&=&{\displaystyle \frac{4\pi\alpha^2}{3s} }~~,\\
\sigma_{\rm res}(s)&=&{\displaystyle\frac{12\pi\Gamma_e^2}{(s-M_{R}^2)^2+M_R^2\Gamma_{\tau}^2} }~~,\\
\sigma_{\rm int}(s)&=&{\displaystyle \frac{8 \alpha \Gamma_e}{\pi \sqrt{s}}
\frac{s-M_{R}^2}{(s-M_{R}^2)^2+M_R^2 \Gamma_{\tau}^2} }~~,
\end{array}
\label{threesxn}
\end{eqnarray}
and
\begin{eqnarray}
F_s(s)=\sigma_{\rm con}(s) \cdot \frac{\pi(3-v^2)}{1-\exp(-\pi\alpha/v)}
\end{eqnarray}
with
\begin{eqnarray}
v^2 \equiv 1-\frac{4m^2_{\tau}}{s}.
\end{eqnarray}
Notice that $v$ is function of $s$ and $m_{\tau}$.  Here $s$ and $m_{\tau}$ are the c.m.~energy and mass of the $\tau$, respectively, and $\Gamma_e$, $\Gamma_{\tau}$ and $M_{R}$, are the di-electron width, the di-tau width, and the mass of the $\psi(2S)$ resonance, respectively.
Finally, we obtain the cross section distributions, according to the above formulas, shown in Fig.~\ref{QEDCalResult}. In the plot, the solid curve indicates the total cross section distribution (black line). The blue, red and pink dashed lines denote the cross section of the continuum, resonance and the interference between resonance and continuum parts, respectively. The continuum plus interference cross section at the center of mass energy 3686.10 MeV with energy spread 1.3 MeV, is calculated to be \mbox{$\sigma_{Q+I}$ = 2.125 nb}.
\begin{figure}[htbp]
\includegraphics[width=10cm]{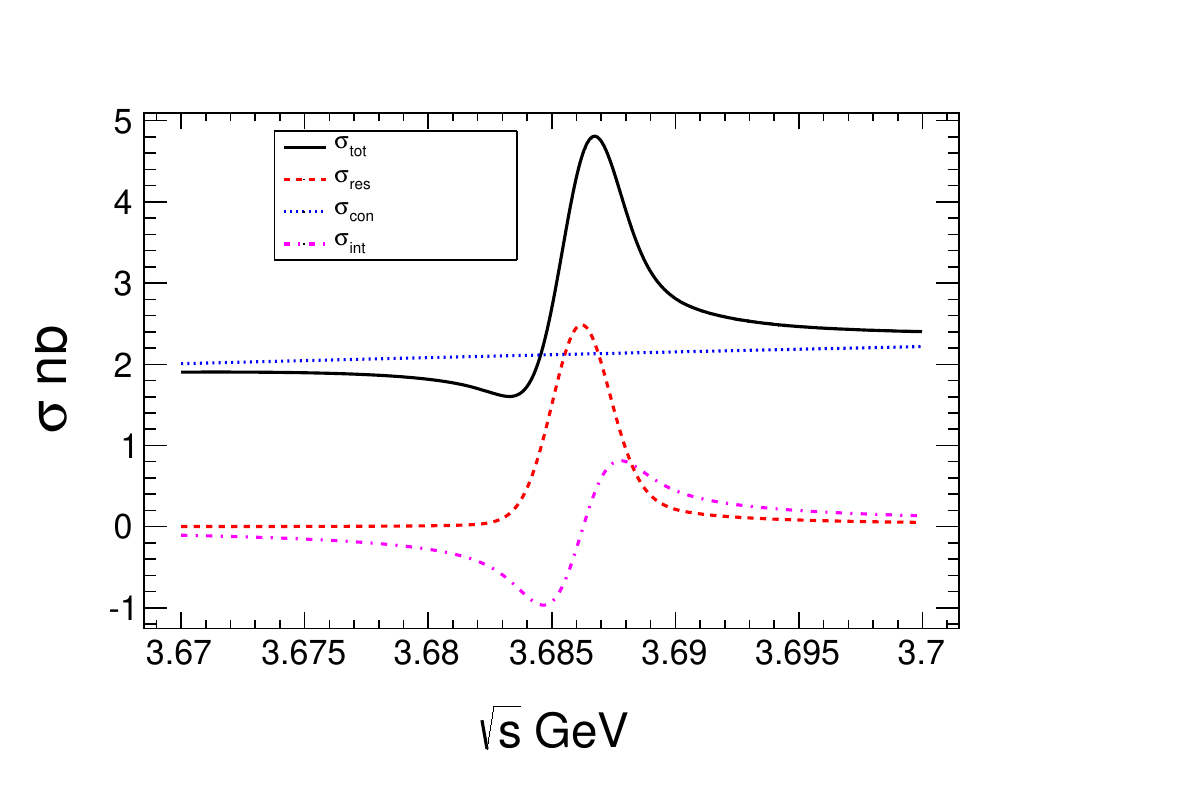}
\caption{The total $e^+e^- \to \tau^+\tau^-$ cross section, and its components, as a function of $\sqrt{s}$ in the vicinity of the $\psi(2S)$ resonance.}
\label{QEDCalResult}
\end{figure}
}


\section{\boldmath Systematic Uncertainty}\label{SysTem}\label{Uncertainty}
{
Systematic uncertainties in the branching fraction measurement come from the luminosity measurement, total number of $\psi(2S)$ events, tracking efficiency, particle identification (PID) efficiency, $\mu^{\pm}$ and $\pi^{\pm}$ separation, number of photons requirement, missing mass~($M_{\rm mis}$) and cos$\theta_{\rm mis}$ requirement, QED cross section calculation, branching fractions of $\tau^+$ decays and background estimation.

\begin{enumerate}
	
    \item The systematic uncertainty in the luminosity measurement is assigned as 1.1$\%$ using the large angle Bhabha and di-gamma events for the $\psi(2S)$ data, where the generator BABAYAGA~\cite{BABAYAGA} is used. Its impact on the measured branching fraction, 1.1$\%$, deduced via error propagation, is assigned as a systematic uncertainty.

    \item The uncertainty in the total number of $\psi(2S)$ events is estimated to be $0.5\% $ from Ref.~\cite{BESIII:2023psipError}.

	\item The uncertainties related to the tracking efficiencies for the electron and muon are estimated to be $0.5\%$ ~\cite{BESIII:2015jmz,BESIII:2018hhz} for each of them, and a total systematic uncertainty of $1.0\%$ is assigned.

	\item To estimate the uncertainties due to the particle identification of the electron and muon, the two processes $\EE\rightarrow\gamma e^{+}e^{-}$ and $\EE\rightarrow\gamma \MM$ are selected as control samples with  $\psi(2S)$ data. For the former, the corresponding MC sample is generated with the BABAYAGA~\cite{BABAYAGA}. And for the latter process, the KKMC generator is used. The difference of the PID efficiency between data and MC for the $e$ and $\mu$ tracks is obtained using a weighted method and is found to be 0.9$\%$ and 0.8$\%$, respectively. Accordingly, 1.2$\%$ is taken as the combined systematic uncertainty from the PID of the electron and muon.

    \item To investigate the uncertainty associated with the requirement applied to distinguish between pions and muons, $\psi(2S)\rightarrow\pi^{+}\pi^{-}J/\psi$ and $\jpsi\rightarrow \rho\pi\rightarrow\pi^{+}\pi^{-}\pi^{0}$ control samples are used, and corresponding MC samples are generated. The maximum relative difference of the detection efficiencies between data and MC simulation obtained with and without this additional requirement is found to be 0.2$\%$. Taking into account the pions background contamination estimated in Sec.~\ref{EvtBkg}, the systematic uncertainty on the branching fraction is calculated to be 1.0$\%$ in the decay $\psi(2S)\rightarrow\tau^{+}\tau^{-}$.

    \item The uncertainty associated with the number of photons is estimated from the efficiencies between data and inclusive MC simulation with and without the photon number requirement, and the difference is found to be 0.2$\%$.

	\item To evaluate the systematic uncertainty associated with the missing mass $M_{\rm mis}$ and cos$\theta_{\rm mis}$, we vary the missing mass from 3.05 to 3.10 GeV$/c^{2}$ or 3.00$/c^{2}$ GeV. The maximum difference of the measured branching fraction, 0.8$\%$, is taken as the systematic uncertainty. Similarly, the relative difference is regarded as the uncertainty by varying cos$\theta_{\rm mis}$ from 0.80 to 0.75 or 0.85, which is approximately 0.1$\%$.

    \item The uncertainty associated with the background contamination is estimated by changing the branching fractions of the corresponding background processes by one standard deviation. The resulting change in the measurement result, 0.3$\%$, is considered as the systematic uncertainty.

    \item Taking into account the uncertainty associated with the c.m. energy measurement, which is 1 MeV from the energy measurement system  and 0.4 MeV from the energy spread conservatively, a resulting difference in the QED cross section calculation, 0.4$\%$, is regarded as the uncertainty.

    \item The uncertainties associated with the branching fractions $\mathcal{B}(\tau^{+} \rightarrow e^{+}\nu_{e}\bar{\nu_{\tau}})$ and $\mathcal{B}(\tau^{+} \rightarrow \mu^{+}\nu_{\mu}\bar{\nu_{\tau}})$ are taken from the PDG~\cite{PDG2022}.

	\item The uncertainty due to the MC statistics is calculated by $\sqrt{(1-\epsilon)/(\epsilon N_{\rm gen})}$ with the number of generated events ($N_{\rm gen}$) and detection efficiency ($\epsilon$), which are $2\times 10^{6}$ and 30.65$\%$, respectively.

\end{enumerate}

\begin{table}[htp]
	\caption{Relative systematic uncertainties in the measurement of the branching fraction of $\psi(2S)\rightarrow\TT$.}
\centering
\begin{tabular}{cc}\\\hline\hline
Source & Uncertainty (\%) \\ \hline
Luminosity & 1.1\\
$N_{\psi(2S)}$ & 0.5\\
Track efficiency	 &1.0\\
PID                & 1.2 \\
$\mu$ and $\pi$ separation & 1.0\\
$N_{\gamma}$ requirement & 0.2 \\
$M_{\rm mis}$ requirement   & 0.8 \\
$\theta_{\rm mis}$ requirement & 0.1\\
Background         & 0.3 \\
Cross section calculation & 0.4 \\
Quoted branching fraction & 0.3\\
MC statistics & 0.1 \\ \hline
Total	             &2.5\\ \hline\hline
\end{tabular}
\label{table:totaluncertainty}
\end{table}

All the systematic certainties are summarized in Table~\ref{table:totaluncertainty}. Combining the individual contributions of all the systematic uncertainties gives a total systemic uncertainty of 2.5$\%$ on the $\psi(2S)\rightarrow\TT$ branching fraction.


\section{\boldmath Calculation of the $\psi(2S)\rightarrow\TT$ branching fraction}\label{BFrCal}
{
To obtain the number of resonant $\tau$-pairs (the signal events), the QED contribution, as well as the contribution from interference, $\sigma_{Q+I}$, is subtracted from the total number of $\TT$ events following Eq.~\ref{BRTTCal}, and $\mathcal{B}_{\TT}$ is calculated by
\begin{eqnarray}
    \mathcal{B}_{\TT}=\frac{\frac{N_{\rm sig}-N_{\rm bkg}}{\mathcal{B}_{e\mu}\cdot\epsilon}-{\sigma_{Q+I}\mathcal{L}}}{N_{\psi(2S)}},
	\label{BRTTCal}
\end{eqnarray}
where $N_{\rm sig}$ is the number of observed signal events, $N_{\rm bkg}$ is the number of background events estimated with the inclusive MC sample and two-photon processes, $\epsilon=30.65\%$ is the detection efficiency estimated from the signal MC sample. $\mathcal{B}_{e\mu}$ is the product of the branching fractions of $\tau^{+}(\tau^{-}) \rightarrow e^{+}\nu_e\bar{\nu_\tau}(e^{-}\bar{\nu_e}\nu_\tau)$ and $\tau^{-}(\tau^{+})\rightarrow\mu^{-}\bar{\nu_\mu}\nu_\tau(\mu^{+}\nu_\mu \bar{\nu_\tau})$ and is equal to 0.06197 quoted from the PDG~\cite{PDG2022}. $N_{\psi(2S)}$ is the total number of $\psi(2S)$ events~\cite{BESIII:2023psipError}, and $\mathcal{L}$ is the luminosity of the data sample~\cite{BESIII:2023psipError}.

The obtained results are summarized in Table~\ref{BrCalTAB}. The branching fraction of $\psi(2S)$ $\to$ $\TT$ is calculated to be
\begin{eqnarray}
\mathcal{B}_{\TT}=(3.240~\pm0.023~\pm~0.081)\times 10^{-3},
\end{eqnarray}
where the first and second uncertainties are statistical and systematic, respectively. The systematic uncertainties are described in Sec~\ref{Uncertainty}.

\begin{table*}[htp]
\caption{Number of observed signal events, number of background events, integrated luminosity, detection efficiency, number of $\psi(2S)$ events, continuum plus interference cross section calculated as described in Sec.~\ref{CrsCal} and the obtained branching fraction of $\psi(2S)\rightarrow\TT$.}
\begin{tabular}{ccccccc}\\\hline\hline
 $N_{\rm obs}$ & $N_{\rm bkg}$ & $\mathcal{L}~(\rm pb^{-1})$ & $\epsilon$ & $N_{\psi(2S)} (10^{6})$ & $\sigma_{Q+I}$ (nb) & $\mathcal{B}_{\TT}(10^{-3})$\\\hline
 280412~$\pm$~530 & 11839~$\pm$~109 & 3208.5 & 0.3065 & 2259.3 & 2.125&3.240~$\pm$~0.023~$\pm$~0.081 \\\hline\hline
\end{tabular}
\label{BrCalTAB}
\end{table*}
}

\section{\boldmath Conclusion}\label{Summary}

In summary, we have measured the branching fraction of $\psi(2S)\to \tau^+\tau^-$ with improved precision. The value of $\mathcal{B}_{\TT}$ together with the world average values of  $\mathcal{B}_{\EE}$ and $\mathcal{B}_{\MM}$~\cite{PDG2022} are reported in Table~\ref{BrEEUUTT}. The comparisons show excellent agreement. Assuming lepton universality, the average value $\mathcal{B}_{l^+l^-}$ is determined to be $(8.06~\pm~0.14 )\times 10^{-3}$. The leptonic width ($\Gamma_{ee}$) of the $\psi(2S)$ has been determined to be $(2.29\pm 0.06)$ keV~\cite{PDG2022}. From the relationship \mbox{$\Gamma_{\rm tot}=\frac{\Gamma_{ee}}{\mathcal{B}_{ll}}$}, the total width of the resonance $\psi(2S)$ is equal to $\Gamma_{\rm tot}=(287~\pm~9)$~keV, which is consistent with the PDG value $(294~\pm~8)$ keV within about one standard deviation.
\begin{table}
\begin{center}
\caption{Leptonic branching fractions of the $\psi(2S)$ in $10^{-3}$}
\begin{tabular}{cccc} \\\hline\hline
		$\mathcal{B}_{\EE}$~\cite{PDG2022} & $\mathcal{B}_{\MM}$~\cite{PDG2022} & $\mathcal{B}_{\TT}$/0.3890 \\ \hline
        7.93~$\pm$~0.17 & 8.00~$\pm$~0.60 & 8.33~$\pm$~0.44 \\ \hline\hline
\end{tabular}
\label{BrEEUUTT}
\end{center}
\end{table}

 This result, along with the previous data of $\Bee$ and $\Bmm$, is in agreement with the relation predicted by the sequential lepton hypothesis.

\acknowledgements

The BESIII Collaboration thanks the staff of BEPCII and the IHEP computing center for their strong support. This work is supported in part by National Key R\&D Program of China under Contracts Nos. 2023YFA1606000, 2020YFA0406300, 2020YFA0406400; National Natural Science Foundation of China (NSFC) under Contracts Nos. 11635010, 11735014, 11935015, 11935016, 11935018, 12025502, 12035009, 12035013, 12061131003, 12192260, 12192261, 12192262, 12192263, 12192264, 12192265, 12221005, 12225509, 12235017, 12361141819; the Chinese Academy of Sciences (CAS) Large-Scale Scientific Facility Program; the CAS Center for Excellence in Particle Physics (CCEPP); Joint Large-Scale Scientific Facility Funds of the NSFC and CAS under Contract No. U1832207; 100 Talents Program of CAS; The Institute of Nuclear and Particle Physics (INPAC) and Shanghai Key Laboratory for Particle Physics and Cosmology; German Research Foundation DFG under Contracts Nos. FOR5327, GRK 2149; Istituto Nazionale di Fisica Nucleare, Italy; Knut and Alice Wallenberg Foundation under Contracts Nos. 2021.0174, 2021.0299; Ministry of Development of Turkey under Contract No. DPT2006K-120470; National Research Foundation of Korea under Contract No. NRF-2022R1A2C1092335; National Science and Technology fund of Mongolia; National Science Research and Innovation Fund (NSRF) via the Program Management Unit for Human Resources \& Institutional Development, Research and Innovation of Thailand under Contracts Nos. B16F640076, B50G670107; Polish National Science Centre under Contract No. 2019/35/O/ST2/02907; Swedish Research Council under Contract No. 2019.04595; The Swedish Foundation for International Cooperation in Research and Higher Education under Contract No. CH2018-7756; U. S. Department of Energy under Contract No. DE-FG02-05ER41374
\bibliography{draft}

\end{document}